# Title: Discovering 'containment': from infants to machines

**Authors:** Shimon Ullman[1]*†, Nimrod Dorfman[1], Daniel Harari[1]

**Affiliations:**
[1]Weizmann Institute of Science (WIS).
*Correspondence to: shimon.ullman@weizmann.ac.il
† Weizmann Institute of Science, Department of Computer Science and Applied Mathematics, 234 Herzl Street, Rehovot 7610001, Israel.

**Rapid developments in the field of automated learning have caused a major shift in the approach to the learning of intelligent systems, from explicit instruction to the automatic learning from a large number of labeled examples. Yet, current methods cannot explain infants' learning, in particular the ability to learn complex concepts without guidance, and the natural order of concepts acquisition. A notable example is the category of 'containers' and the notion of 'containment', one of the earliest spatial relations to be learned [1–5], starting already at 2.5 months [2,5], and preceding other common relations (e.g., 'support' [6]). Such spontaneous unsupervised learning stands in contrast with current highly successful computational models, which learn in a supervised manner, using large data sets of labeled examples [7,8]. How can meaningful concepts be learned without guidance, and what determines the trajectory of infant learning, making some notions appear consistently earlier than others? We present a model, which explains infants' capacity of learning the complex concept of 'containment' and related concepts by just looking, together with their empirical development trajectory. Learning occurs fast and without guidance, relying only on perceptual processes that are present in the first months of life. Instead of labeled training examples, the system provides its own internal supervision to guide the learning process. We show how the detection of so-called 'paradoxical occlusion' provides internal supervision, which guides the system to gradually acquire a range of useful 'containment'-related concepts. The mechanism of using internal implicit supervision is likely to have broad application in other cognitive domains as well as artificial intelligent systems, because it alleviates the need for supplying extensive external supervision, and it can guide the learning process to extract concepts that are meaningful to the observer, even if they are not by themselves obvious, or salient in the input.**

Here we describe a model that learns 'containment', one of the earliest spatial relations to be learned [1–5], and a range of related notions (such as 'support' [6]), fast and without supervision, relying only on perceptual processes that are present in the first months of life. The model goes naturally through stages which appear in infant learning, recognizing first dynamic occlusion events, and then generalizing to static images (Fig. 1). It distinguishes between 'behind', 'in-front' and 'inside' relations, and can tell apart 'tight' and 'loose' fit [4,6,9]. Learning 'support' relations [2,6,10] is more difficult in the model and emerges only later (Fig. 1f, 2f). The model deals with related concepts (e.g. 'cover' [2,5]) and predicts developmental steps that can be tested empirically. The focus of the model is on learning from visual experience, which normally plays a major role in the first months of life.

The perceptual capacities included in the model prior to learning 'containment' are (i) segmentation of a moving region from its surround (*Methods C1, C2*), and (ii) identification of motion discontinuities, and using them to locate boundaries and assign boundary ownership (which side belongs to the object [11,12], *Methods C3*). The model is then exposed to videos and images and acquires on its own and in a natural order a sequence of concepts related to containment.

In both dynamic and static visual input, containment is identified in the model as an instance of 'paradoxical occlusion', defined as a situation where an object *O*, which occludes a second object *C,* is at the same time also occluded by *C* (Supplementary Fig. 1b, 1c). It is known that during early visual experience infants develop the ability to segregate a scene into distinct objects, and determine occlusion relations between them [13–15]. Our model suggests that the occurrence of an unusual paradoxical occlusion event is noted by the system and serves as an early internal signal for containment configurations, which then get elaborated by additional learning. The model demonstrates that paradoxical occlusion serves as an efficient internal guidance, which leads to the learning of containment and related notions in a human-like manner, and based on similar initial capacities.

Briefly, the model proceeds along the following stages (Fig. 2, *Methods, Stages of containment recognition*). In the dynamic case, when an object *O* is inserted into a container *C*, at the moment of a containment event (entering a cavity in *C*), *O* turns from progressively occluding *C* to become occluded by it, signaling a paradoxical occlusion (Supplementary Fig. 1b). This event (detected by a switch in boundary ownership), combined with simple region segmentation (the switch occurs inside region *C*), identifies unambiguously the container and contained object (Fig. 2a, *Methods C1-C3*). In our experiments, following familiarization, the model implementation correctly identified 91% of the test dynamic events (distinguishing 'containment' from 'occlusion', *Materials, Dynamic test sequences*).

In evolving from the dynamic to the static case, the model uses motion discontinuities to extract object boundaries [12,16], which become a part of the object representation. The ability to segregate object-regions in images from their background is consistent with the known capacities of infant at the stage of learning static containment [15,17–20]. At this stage, the model determines occlusion relations based on detected object boundaries instead of relying exclusively on motion cues. Object boundaries are learned from motion discontinuities, which include the external object boundaries, and, for a container, also the internal boundary γ (at the container's rim), which is an inherent part of a container (Fig. 2b, *Methods C4, C5*). If an object *O* is contained in a container *C*, it is detected as occluding the container *C*, and at the same time occluded by it at the internal boundary γ, signaling again a paradoxical occlusion (Fig. 2b, Supplementary Fig. 1c, *Methods, Stages of containment recognition*). At this stage, the model implementation correctly identified 89% of the static test images (distinguishing 'contained', 'in-front', 'behind'; *Materials, Static test images,* low-view images).

A distinction that follows in the model is between 'tight' and 'loose' containment relations, which infants are sensitive to [4,9,21]: the containment is tight if β, the boundary between the object and container, and the rim γ are similar in size, and loose if β is significantly smaller (Fig. 2c,

*Methods, Stages of containment recognition*). The judgments of 'loose' and 'tight' produced by the model also show similarity to adult humans' judgments from the same test images (r=0.71, *Supplementary Methods*). A number of additional relations are learned by the model in an increasing order of complexity. High-view containment is more difficult in the model and requires additional learning (distinguishing 'front' from 'back' object regions, *Methods C6*), because the object *O* is no longer adjacent to the boundary γ and is not occluded by *C* (Fig. 1d, Supplementary Fig. 1d, *Methods, Stages of containment recognition*). The model at this stage identified correctly 82% of both low and high-view static test images (*Materials, Static test images*).The 'support' ('on-top') relation (Fig. 1f, 2f) is more difficult still, since the discontinuity boundary γ is replaced in this case by a convex object edge, which is more difficult to detect by motion discontinuities [22] (*Supplementary Methods, Later stages)*). Finally, a 'cover' relation [2,5] (Fig. 1e, 2e) is similar to containment (in both, object *O* is partially inserted into a cavity in *C*), and will be learned in a similar way. However, this learning will depend crucially on whether the internal discontinuity γ is made visible during familiarization (Supplementary Fig. 2). The model predicts that low-view 'containment', high-view, and 'support' will be acquired in this order, and that 'cover' will be learned spontaneously, provided that the rim γ will be visible during familiarization, but will not be learned otherwise at this stage (*Methods, Stages of containment recognition)*.

Concepts related to spatial relations in general and containment in particular are a fundamental component of human cognition, and they play a useful role in reasoning about a broad range of physical phenomena [23,24]. The model shows how containment concepts can emerge early and without explicit supervision, and in a predictable order. The main mechanism that allows this learning is the detection of paradoxical occlusion, and its use for guiding the learning process.

The ability to detect and pay attention to paradoxical occlusion can be expected in early developmental stages, when infants rapidly learns to detect object boundaries and establish occlusion relations [25,26]. This signal then provides internal implicit supervision, and guides the system to gradually acquire a range of useful containment-related concepts.

The sensitivity to paradoxical occlusion may be a special case of violation-of-expectation, since unlike simple occlusion, in paradoxical occlusion two opposing occlusion relations exist between the same two objects. Based on general developmental processes [27], this can enhance the learning of containment relations and their implications.

Detecting static paradoxical occlusion may be aided by depth information [19,28] however, binocular vision [29] and pictorial depth perception [30] evolve gradually starting at a few months of age, and their contribution to early stages of containment learning is likely to be limited. The model focuses on early stages of learning to identify containers and containment. Reaching a comprehensive understanding of concepts related to 'containment' at an adult level [23] is likely to develop over an extended period, and to incorporate non-visual components, including sensory-motor manipulation.

The mechanism of using internal implicit supervision to guide learning is likely to have broader application in other cognitive domains, because it serves two highly useful and general roles. First, it alleviates the need for supplying extensive external supervision, and second, it can guide the learning process to extract concepts that are meaningful to the observer, even if they are not by themselves highly salient in the visual input. Such aspects of cognitive learning discovered in infants can conceivably be adapted for use by future machine learning systems, which currently often rely on large annotated data sets supplying external supervision [7,8]. An intriguing alternative is to develop machine learning methods, which use prolonged unsupervised training to discover on their own useful guiding signals, which can then support fast unsupervised learning from experience.

# Figures

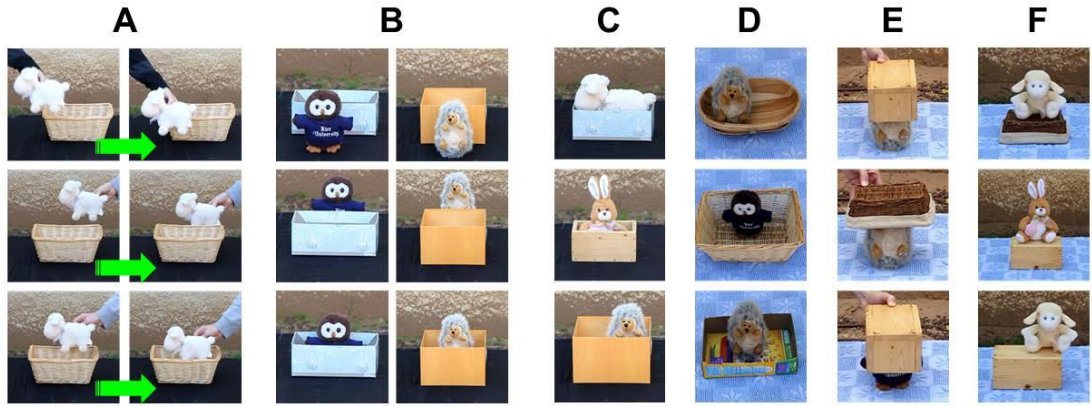

**Figure 1: 'Containment' developmental stages**. **(A)** Dynamic input. Short temporal sequences depicting (top to bottom): 'in-front', 'behind' and 'inside' events. **(B)** Static input. Single-frame images of (top to bottom): 'in-front', 'behind' and 'inside' relations. **(C)** 'Tight' (top) and 'loose' (middle, bottom) fit. **(D)** High-angle view (contained object is not occluded by the container). **(E)** 'Cover' relations. **(F)** 'Support' relations.

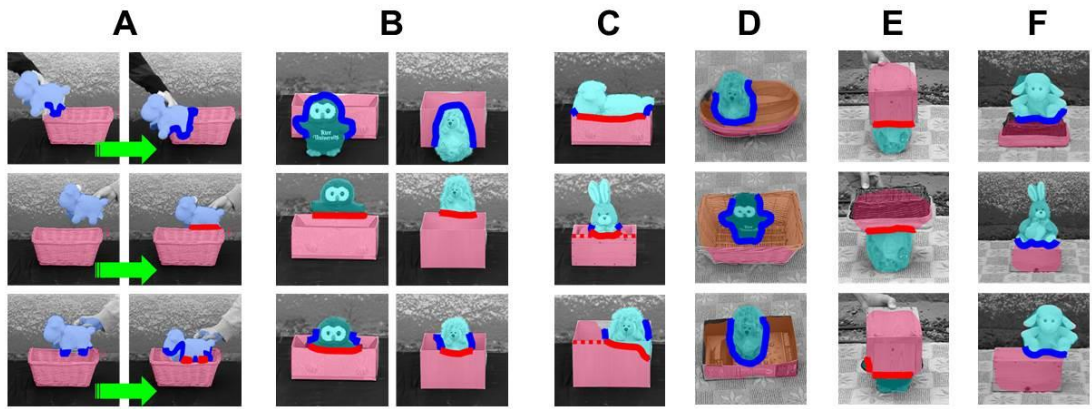

**Figure 2. Schematic illustration of the computations used by the model. (A-D)** Learning to identify 'containment' relations by 'paradoxical occlusion' (bottom row), 'in-front' (top), and 'behind' (middle). Boundaries marked in blue are owned by a simple object and in red by a container. **(A)** Identifying containment in dynamic input. Each row represents a dynamic event depicting an object placed (top to bottom) 'in-front', 'behind', or 'inside' a stationary object. The model segments the objects (colored regions), detects the motion boundary between them, and detects the switch from 'blue in-front of red' to 'red in-front of blue'. **(B)** Detecting 'containment' in static images (two examples, bottom). The internal boundary at the rim is used to identify a 'paradoxical occlusion', where the object is in front of the container (at the blue boundary) but behind it at the rim (red boundary). **(C)** 'Loose' vs. 'tight' fit is measured by the fraction of the detected boundary (solid-red) relative to the full length of the internal boundary (dotted-red). **(D)** High-angle view: The container's region is segregated into 'front' and 'back' regions (shown here in orange), at the internal boundary (*Methods C6, A9-A10*). Detection of 'containment' is extended to include occlusion (blue boundary) confined to the 'back' region (*Methods A9, Supplementary Methods SA9*). **(E)** 'Cover' relation (when the internal part is invisible) and 'support' (in **F**) are related to containment but in the model they require additional learning and predicted to appear at later stages (*Methods, Stages of containment recognition, Supplementary Methods*).

# Methods:

**Data for the computational processing.** Data consisted of familiarization video sequences, and test data including both videos and static images. The familiarization data were used to introduce the participating objects to the model. The test data were used to evaluate the model's performance in detecting containment along with related relations ('behind', 'in-front', 'on-top', Supplementary Fig. 3).

The videos and images (640×360 pixels) were taken with a stationary camera from two viewpoints: a low viewing angle, where objects were partially occluded when placed inside a container, and a high viewing angle, where objects were fully visible inside a container (Fig. 1a-d, Supplementary Fig. 3). Objects used in the experiments included seven containers (wooden box, baskets, card boxes), and five non-container objects, termed 'simple' objects (stuffed animals). For support ('on-top') and cover relations, we used two of the containers, but with their open side facing down Fig. 1e, 1f).

*Data for familiarization sequences.* Short video sequences were presented to the model prior to testing, and were used by the model to learn about the participating objects, by detecting object boundaries and separating the objects from the background. Similar familiarization episodes are used routinely in infant experiments for introducing the participating objects to the infants [31,32]. Each video introduced a single object being moved by a hand against a static background. We also used an image of the object at rest, without a holding hand, to separate the hand from the object. The motion included some wiggling, which made the motion boundary at the container rim easier to detect (Supplementary Fig. 2). For each object there were 2-4 sequences, with a total duration of 2 seconds for simple objects and 8 seconds for containers.

*Data for dynamic test sequences.* Short video sequences were used for testing the automatic detection of dynamic containment events ('inside') and distinguish them from occlusion events ('in-front' or 'behind'). Each video depicted a stationary container and a moving object being placed inside, in-front, or behind the container. The videos were taken from a low-viewing angle (where contained objects are occluded by the containers at containment events, Fig. 1b). There were a total of 176 test video sequences: 59 inside, 57 in front, 60 behind with various object-container pairs. Each video was 1 second long.

*Data for static test images.* Single-frame images were used for testing the detection of different spatial relations between objects in a static setting. Each test image showed a simple object 'inside', 'in-front' or 'behind' a container. Test images included both low (176 images) and high viewing angles (175 images). There were 351 static test images in total, for different object-container pairs, different spatial relations and the two viewing angles (Supplementary Fig. 3).

**Computational model.** The model includes a number of capacities that develop over time (*Methods, Capacities of the model C1-C6*). It is equipped initially with capacities (*C1-C3*), which exist prior to learning; additional capacities are acquired in subsequent stages in a dependent manner. These capacities are processes dealing with object segregation and boundary detection, starting from moving objects and generalizing to static scenes. We also describe in this section how these capacities are implemented in the model (Supplementary Fig. 4, *Methods, Algorithmic implementations A1-A10*), and how they are used to acquire 'containment' concepts in a sequence of stages (Fig. 1, *Methods*, *Stages of recognition*). The capacities and algorithms used at the different stages are summarized in Supplementary Table 1.

*Capacities of the model*
 (*C1*) *Figure-ground segmentation of moving regions*. The moving region is separated from the background based on its motion, and the model constructs a simple representation of the region.
(C2) *Detection of familiar regions in a static scene*. Based on the representation in (*C1*), the model can detect a familiar region and separate it from its background in a static scene (Supplementary Fig. 4a).
(*C3*) *Detection of boundaries and their ownership at motion discontinuities*. The model detects object boundaries by motion discontinuities, and determines 'ownership' direction (object side).
(*C4*) *Detection of internal boundaries of familiar moving regions*. Based on the representation in (*C1-C3*) above, the model can detect motion discontinuities within the region of an object as internal boundaries (typically produced at a container's rim). The model represents the internal boundary as a part of the object.
(*C5*) *Detection of familiar boundaries in a static scene*. In a container, based on the representation in (C3, *C4*), the external and internal boundaries can be detected in a static scene (Supplementary Fig. 4b).
(*C6*) *Detection of familiar internal regions in a static scene*. In a container, based on (*C2-C5*), the model can discriminate between the 'front' and 'back' sides of the internal boundary (Supplementary Fig. 4c).

*Stages of containment recognition*
*Recognizing dynamic containment*. For recognizing containment relations between objects (as well as 'in-front', 'behind') in dynamic displays, the model is first presented with brief familiarization video sequences of moving objects (both containers and non-containers), from which it learns to detect the objects' regions in subsequent test videos (*Methods C1, C2, A1, A2, Supplementary Methods*).
In dynamic sequences, during a containment event, when an object $O$ is inserted into a container $C$ (entering a cavity in $C$), object $O$ turns from progressively occluding $C$ to become partly occluded by $C$, signaling a paradoxical occlusion (Supplementary Fig. 1b). This event is detected by a switch in boundary ownership (*Methods C3, A3*): when this switch occurs inside region $C$ rather that at its boundary, it identifies unambiguously the container $C$ and the contained object $O$ (Fig. 2a, Supplementary Fig. 1b, *Methods A7*).
*Recognizing static containment*. For recognizing static containment between objects (as well as 'in-front', 'behind'), the model moves to its next stage. The development is the addition of boundaries, including the internal boundary at the container's rim, to the object model (Supplementary Fig. 1c, 4, *Methods C4, C5, A4, A5*). At this stage, containers and non-containers produce different object-models during the familiarization stage, since container regions include an internal boundary (at the container's rim), which is not present in non-container regions. In a static scene, paradoxical occlusion is signaled not by a dynamic switch, but by conflicting occlusion relations between the two objects. The conflicting cues arise along different parts of the common boundary between the container $C$ and the object $O$ inside its cavity (Supplementary Fig. 1c). A part of the common boundary is owned by the object $O$, and another part, along $C$'s internal boundary, is owned by the container $C$. (Fig. 2b, Supplementary Fig. 4, *Methods A8*). Similar to the dynamic case, the container $C$ both occludes and is being occluded by $O$.
*'Tight' versus 'loose' fit*. In a containment event, an object $O$ can either fit tightly inside a container $C$, or it may occupy only a part of $C$'s cavity, and may be free to move within it (Fig. 1c). Discrimination between tight and loose static containment is based in the model on the internal boundary of the container (*Methods C5*) and the object regions on its two sides (*Methods C1-C4*), extracted automatically during the familiarization stage. The model produces a measure of the containment 'tightness' based on the proportion between the total length of *C*'s internal boundary, and the length of the common part of the internal boundary shared between *C* and the inserted object *O*. This stage depends, therefore, on the reliable detection of C's internal boundary through most of its length (Fig. 2c, *Methods A9*).

*Recognizing static containment from a high view*. For recognizing static containment between objects from a high view (Fig. 1d), an additional capacity is required. In the early stage, an object in the image is represented by a single region, (*Methods C1-C4, A1-A4*). Subsequently, internal object boundaries are detected (*Methods C4, C5, A4, A5*). The addition of an internal boundary naturally breaks the single object region into two regions joined along the internal boundary. For dealing with high-view angle, the model uses the representation of the two sub-regions of a container (simple objects, without an internal discontinuity, are still represented by a single region). We term the two sub-regions 'front' and 'back', based on the detected border ownership (*C6,* the owner of the internal boundary is the 'front'). High view containment is detected when all the common borders between *O* and *C* are owned by *O*, and separate O from the 'back' region of *C* (*Methods A10*).

*Later stages: recognizing support and cover relations.* Our simulations show that an additional capacity is required for recognizing a support relation (i.e. that an object is 'on-top' another object, Fig. 1f, 2f). Briefly, the surface boundaries of a supporting object have no depth discontinuity (only surface orientation), making them more difficult to detect [22,33] (*Supplementary Methods*).

A cover relation [2,5] (Fig. 1e, 2e) is similar to containment (in both, object *O* is partially inserted into a cavity in *C*), and will be learned in a similar way. However, this learning in the model will depend crucially on whether the internal discontinuity at the covering object's opening rim is made visible during training (Supplementary Fig. 2).

*Algorithmic implementations*

Described below are the algorithms implementing the model's capacities and processes. Algorithms for motion computations, segmentation and regions representations (*Methods A1-A6*) use primarily existing computer vision schemes, extended by algorithms dealing with containers and containment (*Methods A7-A10*).

(*A1*) *Figure-ground segmentation of moving regions*. The model applies a standard optical flow computation [34,35] between each pair of successive video frames in the input sequences to evaluate the motion in the scene. Moving regions are separated from the background similar to [36–39], and their boundaries are identified as loci of discontinuity (or sharp transition) in image motion [40] (Supplementary Fig. 2). Local motion discontinuities are detected using directional gradients of the optical flow [41,42]. Internal boundaries between figure sub-regions do not participate in the segmentation process at this stage. The model produces a representation of the moving figure region. The object representation is a so-called star model [43], which is a configuration of local regions surrounding a common center. Each region includes a description of its appearance [44] its segmentation mask, and offset from the object center [40,43,45,46] (*Supplementary Methods*).

(*A2*) *Detection of familiar regions in a static scene*. We use an existing model for representing an image region (in *Methods A1*), and detecting a similar region in a new image [40,43,45]. The model is tolerant to occlusion and moderate scale changes, and can robustly detect partially occluded objects when located in- front, inside or behind other objects in the scene.

(*A3*) *Detection of boundary ownership at motion discontinuities*. We use motion cues to assign, for each pixel along a motion discontinuity, the direction of the region that 'owns' the boundary. The owner region is defined as the neighboring region that moves together with the boundary [16]. The algorithm computes the image motion on the two sides of the motion boundary, denoted by $V_1$, $V_2$. It also tracks the displacement of the motion boundary itself, $V_b$. Ideally, the velocity of the owning region should match the measured velocity of the boundary (since the boundary is the edge of the moving region, Supplementary Fig. 5). The algorithm therefore computes $||V_1-V_b||$, $||V_2-V_b||$, and the owner is identified by the side that produces the smaller magnitude; the magnitude itself is used as a confidence score.

(*A4*) *Detection of internal boundaries of familiar moving regions*. Based on (*Methods A1, A2*), the object is separated from the background along its external boundaries. The internal boundary at a container's rim is detected as a motion discontinuity boundary during the dynamic familiarization stage (*Methods A3*). Since this motion discontinuity may be noisy and discontinuous, noise is reduced using consistency of boundary ownership direction along the detected discontinuities (*Supplementary Methods*). The model produces a representation of the internal boundary similar to the representation of external boundaries in (*Methods A1, A2*) (the internal boundary separates between two sub-regions of the object rather than between the object and background).

(*A5*) *Detection of familiar internal boundaries in a static scene*. Based on (*Methods A4*) the detection of the internal boundary separating two object sub-regions in a static scene, uses the same algorithm applied to locate external boundaries, separating figure from background regions (*Methods A1, A2*, *Supplementary Methods*).

(*A6*) *Detection of familiar internal regions in a static scene*. The internal boundary along the rim naturally divides the container's region into two sub-regions, 'front' and 'back' Fig. 2d, Supplementary Fig. 4c). Based on (*Methods A3, A4*), during the familiarization phase, the model discriminates between the front and back regions of a container separated by an internal boundary at the container's rim (Supplementary Fig. 2d). The two sub-regions become a part of the object model. Using this representation, the model uses the algorithm used to detect regions and boundaries in a static image (*Methods A1, A2, A4, A5*) to also detect the front and back regions in a static scene (*Supplementary Methods*).

(*A7*) *Recognizing dynamic containment*. The algorithm tracks the moving boundary between two object regions in the image, $R_1$, $R_2$, which have been learned during the familiarization period. The algorithm computes boundary ownership at each pixel along the moving boundary (Supplementary Fig. 1b); since the ownership measurements can be noisy, they are integrated over 200 milliseconds, and the direction of ownership is determined by majority vote. The model detects a paradoxical occlusion (a containment relation) when ownership direction switches from $R_1$ to $R_2$ between two consecutive time windows. Furthermore, this switch signals that the final owner, R2, is the container (*Supplementary Methods*).

(*A8*) *Recognizing static containment*. Static containment is detected by paradoxical occlusion: the simple object *O* occludes the container *C*, and it is also occluded by *C* at the internal boundary (Fig. 1b, Supplementary Fig. 1b). Since the internal boundary is partially occluded on one side by the inserted object, the detection algorithm is applied separately to the two sides of the boundary. Paradoxical occlusion is detected based on the simultaneous detection of opposing occlusion relations along the common borders between *O* and *C* (*Methods A2, A4-A6*, *Supplementary Methods*).

(*A9*) *Measuring containment 'tightness'*. In a containment event, the model extracts the entire length of the internal boundary of the container *C* and the length of the common part of the internal boundary shared between *C* and the inserted object *O*. The model then produces a measure of the containment 'tightness' based on the proportion between the two lengths. A measure close to 1 (similar lengths) indicates a 'tight' fit, and a high measure (the common boundary between *C* and *O* is small compared with the full internal boundary of *C*) indicates a 'loose' fit.

(*A10*) *Recognizing static containment from a high view*. In a high view, the object is entirely surrounded by the container's 'back' region (Supplementary Fig. 1d). Based on (*Methods A1, A2, A4-A6*) the model detects the objects' regions and boundaries in the static scene, including the container's front and back regions. The model assigns a boundary ownership to all the common borders between the simple object *O* and the container *C*. The model detects high-view containment when all the common borders between *O* and *C* are owned by *O*, and separate O from the 'back' region of *C*. By analyzing the local ownership along the common borders and the 'front'/'back' regions, the model separates containment from 'in-front' and 'behind' relations (*Supplementary Methods*). The model does not require supervision to learn high-view containment, but finds this new relation based on clustering of occlusion relations (*Supplementary Methods, SA10).*

**Acknowledgments** We thank Chen Yu and Maryam Vaziri-Pashkam for discussions and comments. The work was supported by European Research Council (ERC) Advanced Grant "Digital Baby" (to S.U.).
**Author contributions** S.U., N.D. and D.H. designed research; D.H. and N.D. performed research; D.H. and N.D. analyzed data; S.U., N.D. and D.H. wrote the paper. S.U., N.D. and D.H. contributed equally to this work. All authors discussed the results and commented on the manuscript.

# Supplementary figures and tables:

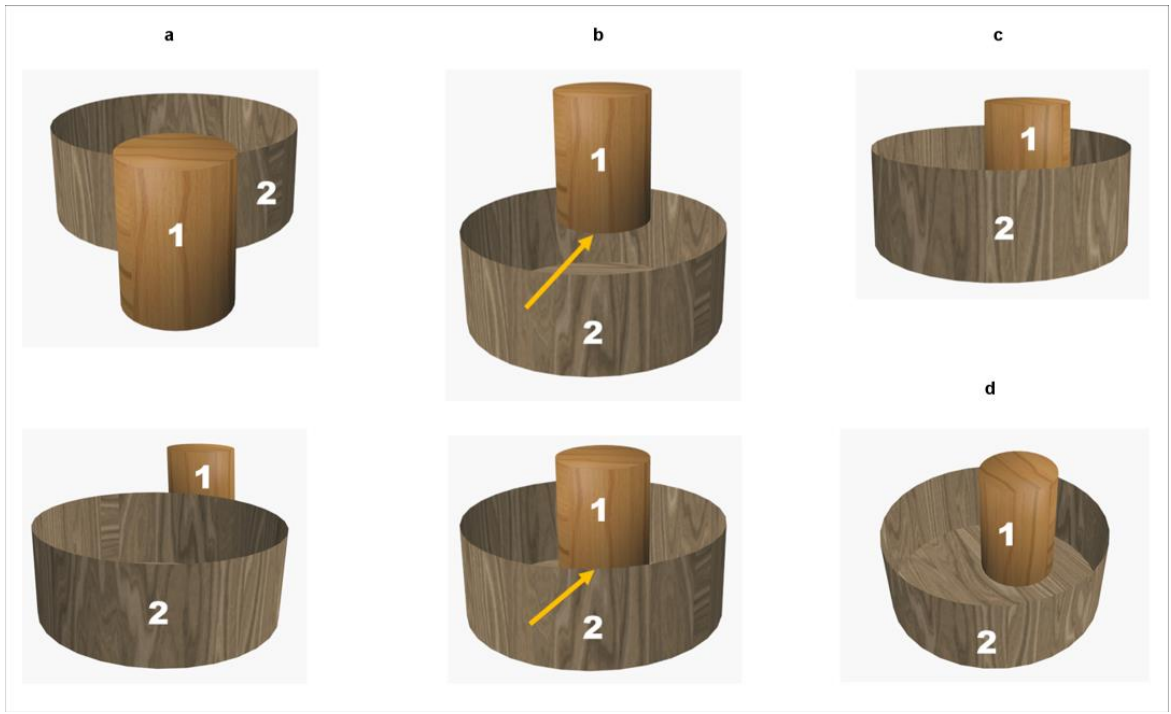

**Supplementary Figure 1: Simple occlusion vs. containment. (A)** Simple occlusions. (Top) An object (1) is in front of, and occludes a second object (2), or (bottom) is behind, being occluded by (2). **(B)** Dynamic containment occurs when a switch in boundary ownership (at orange arrow) between (1) and (2) signals a 'paradoxical occlusion'. **(C)** Static containment at a low view is detected as a 'paradoxical occlusion', where (1) occludes (2) and also being occluded by (2) at the same time. **(D)** Static containment at a high view: the common boundary is owned by (1) and is between (1) and the 'back' region of (2).

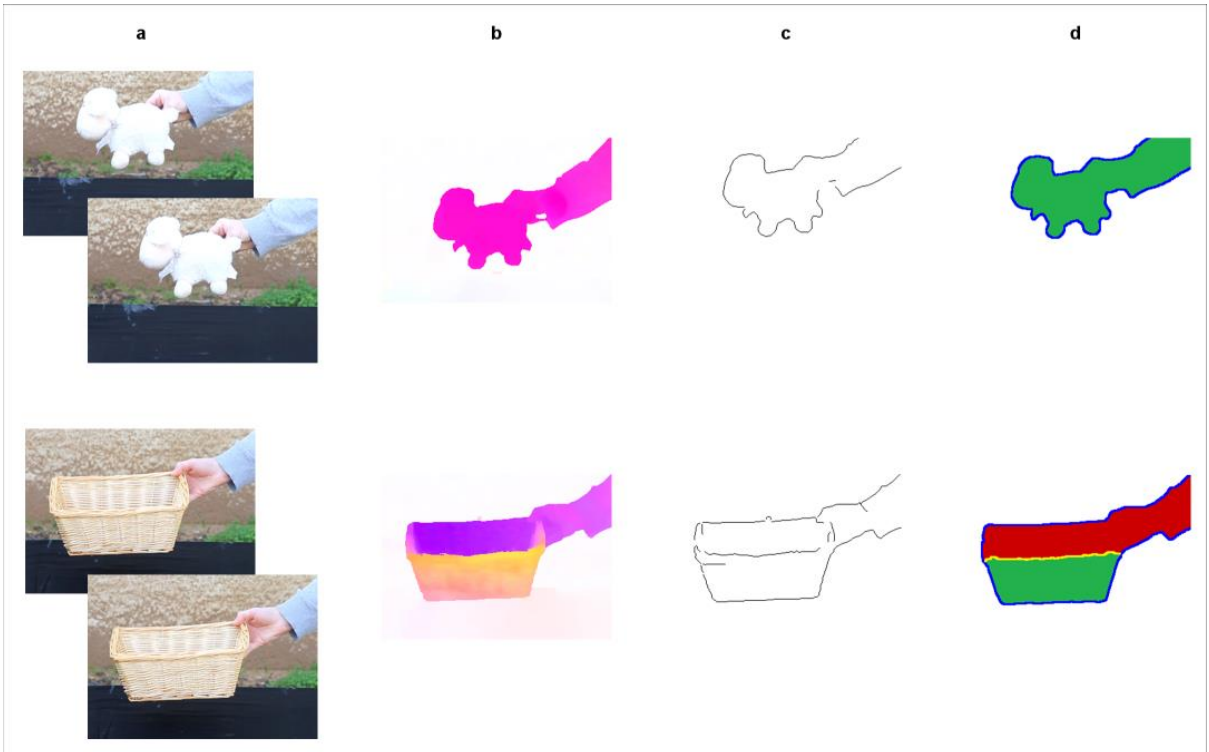

**Supplementary Figure 2: Figure-ground segmentation**. Analysis steps for a simple, non-container object (top) and a container (bottom). **(A)** Two consecutive frames from a familiarization video sequence. **(B)** Computed optical flow between the two frames using [35]. Direction and magnitude are represented by hue and saturation respectively **(C)** Motion discontinuities are computed from local gradients of the optical flow. **(D)** Figure-ground segmentation for the two types of objects: (top) a simple object is separated from the background by an external boundary; (bottom) a container has in addition two sub-regions separated by an internal boundary at the container's rim. Later, the hand is separated from the object using an additional image of the object at rest without the hand holding it.

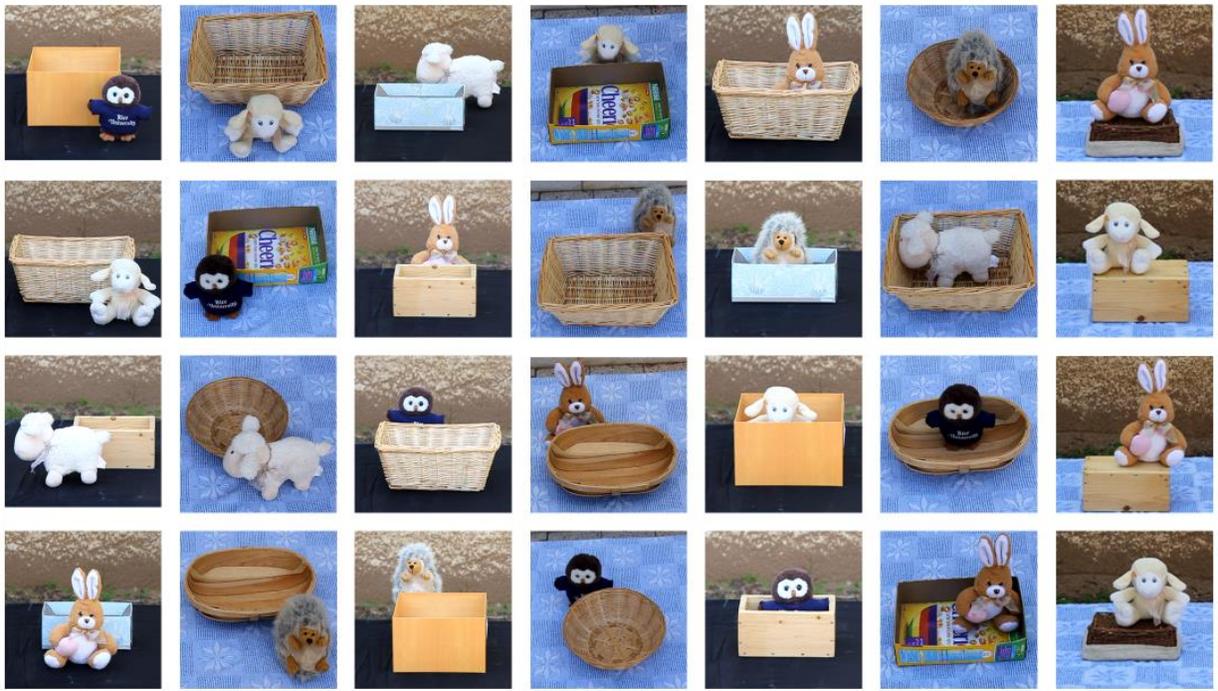

**Supplementary Figure 3: Examples from the dataset.** Examples from the test images depicting various objects (staffed animals) at different spatial relations ('in front', 'behind', 'inside', 'on-top'), with various containers, and at multiple views.

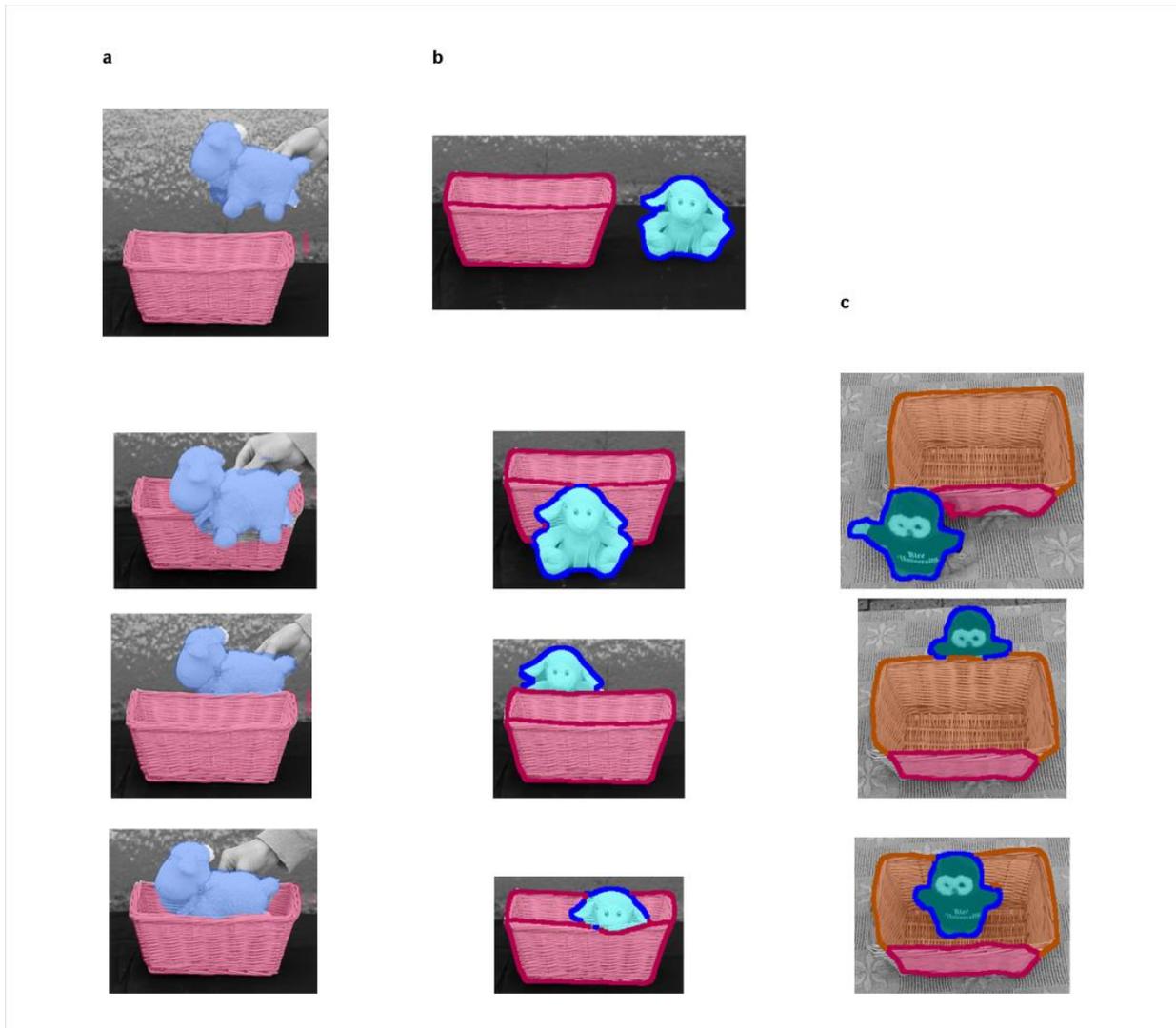

**Supplementary Figure 4: Detection capacities of the model.** (**A**) Detection of familiar regions (*Methods C1, C2*) - containers (red region) and simple objects (blue region). (**B**) Detection of boundaries (*Methods C4, C5*) - external in simple objects (blue lines) and both external and internal (at the rim) in containers (red lines). (**C**) Detection of container's sub-regions (*Methods C6*) - 'front' (red region) and 'back' (orange region) separated by internal boundaries at the container's rim (the simple object is in blue).

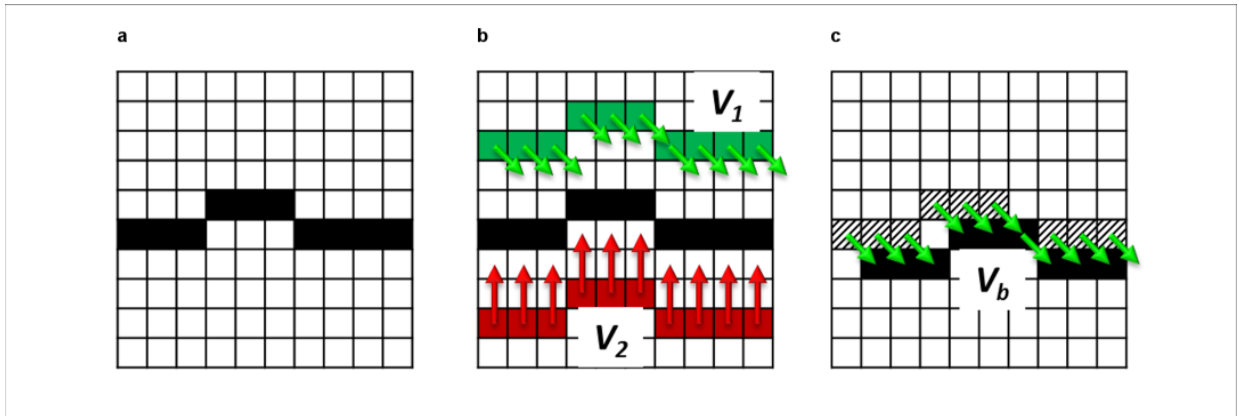

**Supplementary Figure 5: Boundary ownership computation.** (**A**) A patch of pixels in the first frame, showing a segment of the motion boundary in black. (**B**) Motion fields on the two sides, along strips parallel to the boundary (bottom strip in red; top strip in green). Arrows show the optical flow $V_1$ and $V_2$ in these strips. (**C**) The motion boundary in the second frame (solid black) has been displaced from its previous position (shaded region) according to the motion $V_b$. This motion of the boundary is more similar to the motion $V_1$ of the top region (green arrows) than to $V_2$, and therefore, owned by the top region (the foreground).

| Recognition stage | Model's capacities | Algorithmic Implementations |
|---|---|---|
| Dynamic containment | C1-C3 | A1-A3, A7 |
| Static containment | C1-C5 | A1-A5, A8 |
| 'Tight' and 'loose' fit | C1-C5 | A1-A5, A8, A9 |
| High-view containment | C1-C6 | A1-A6, A10 |

**Supplementary Table 1: A summary of recognition stages, model's capacities and algorithmic implementations.** Shown for each stage are the capacities included in the model (*Methods, Capacities of the model*) and the algorithmic implementations (*Methods, Algorithmic implementations*).

## Supplementary Methods:

*Stages of containment recognition*

*Tight versus loose containment*. 5 human subjects were asked to judge between 'tight' and 'loose' fit in static test images. The judgments of 'loose' and 'tight' produced by the model were correlated with the human judgements (r=0.71). This shows that the manner in which the algorithm extracts and uses boundary information from visual cues is correlated with humans' tight/loose visual judgement.

*Later stages: recognizing support and cover relations.* The detection of object boundaries at motion discontinuities along the container's rim is essential for recognizing a containment relation. Similarly, the detection of object boundaries on the surface of a supporting object is essential for recognizing a support relation. These two types of boundaries arise from different kinds of discontinuity [33]: contours of depth discontinuity at the container rim (occlusion contours), vs. contours of discontinuity in surface orientation, e.g. along the internal edges of a box (or even no discontinuity). We tested computationally which type of contours produce stronger motion discontinuities. For this purpose, we compared the detection confidence score (*Methods A3, A4*) of motion discontinuities along edges of a box, that could be convex (solid boxes) or concave (open boxes).

We simulated short video sequences of rotating artificial 3D cubes (both concave and convex) with 10 different textures. The mean confidence score for the concave (container) cubes (2.37±1.76) was significantly higher (two-sample tailed t-test: $p < 10^{-6}$) than the mean confidence score for the convex (closed box) cubes (0.28±0.64). Due to the crucial role of the boundaries, this makes the detection of support relation more difficult compared with containment relations.

*Algorithmic implementation*

(*SA1*) *Figure-ground segmentation of moving regions*. The object representation is a so-called star model [43], which is a configuration of local regions surrounding a common center. The object center is determined at the first familiarization video frame as the center of mass of object pixels. In each consecutive frame, the center is found by detecting the object using the model representation from previous frames.

Object models are learned from videos where a grasping hand was moving the object. In order to separate the hand from the learned object model, we detect the object once in a reference image that shows the object without a hand. The detected object region defines the valid offsets from the object center. Any object descriptors in the model that exceed these offsets by more than 20 pixels are discarded.

(*SA2*) *Detection of familiar regions in a static scene*. Given a static image, local appearance descriptors [44] are densely extracted throughout the image. For each descriptor, the model retrieves the *k* nearest neighbors (*k*-NN) from the learned object model. Each neighbor votes for the location of the object center with a relative weight proportional to its learned predictive power. The image location with the highest total votes is detected as the object center [43]. To segment out the object region (and subsequently its boundaries), the model projects back the figure-ground masks associated with the image features at the corresponding relative offset from the detected center [40,45]. The model's segmentation capability develops with the different stages stating with a single object foreground region in the first stage, and including the object external and internal boundaries, and 'front' and 'back' regions at later stages. The model is tolerant to object scale differences of ±10% with respect to the familiarization scale, and therefore can robustly detect objects that appear closer or further away relative to their distance from the camera during familiarization, when located in-front, inside or behind other objects in the scene.

(*SA3*) *Detection of boundary ownership at motion discontinuities*. The local confidence scores computed for determining the boundary owner are integrated along small boundary segments (Kovesi P., MATLAB and Octave Functions for Computer Vision and Image Processing, http://www.peterkovesi.com/matlabfns/, 2000).

(*SA4*) *Detection of internal boundaries of familiar moving regions*. During the familiarization phase, the model is presented with brief video sequences of moving objects (containers and non-containers). The model segments the object from the background, while preserving internal regions separated by internal boundaries (where part of the object occludes other parts of the object). In particular, the model produces the segmentation from boundaries similar to [47,48], but where the boundaries are motion discontinuities rather than intensity edges. The method proceeds by producing a so-called 'over-segmentation', which consists of a relatively large number of small regions that are compatible with the detected boundaries. Next, regions separated by low boundary ownership measure (*Methods*

*A3*) are iteratively merged until there are two regions left. A high discontinuity measure between the last two regions indicates that the object consists of 'front' and 'back' regions separated by an internal boundary.

(*SA7*) *Recognizing dynamic containment*. To increase noise robustness, the final detection also requires a minimal difference between the fraction of the common border owned by each object (set empirically to 0.4), and a minimal length of assigned boundary (set to 30 pixels) on average per frame.

(*SA8*) *Recognizing static containment*. Static containment is based on the detection of opposing occlusion relations along the common borders between the object *O* and container *C*. Decision was based on a measure of the paradoxical relation defined by:

$$Paradoxical\ ratio = \frac{min(o,c)}{o+c}$$

where *o*, *c* are the number of border pixels owned by *O* and *C* respectively. Threshold was set empirically to 0.2 and was fixed for all experiments, yielding accuracy of 89.2% correct decision (results were insensitive to the threshold in the range 0.1-0.5).

(*SA10*) *Recognizing static containment from a high view*. There are four possible local occlusion relations between the object *O* and the container's (*C*) front and back regions: *O* occludes *C*'s front region, *O* occludes *C*'s back region, *C*'s front region occludes *O* and *C*'s back region occludes *O*. The model uses a normalized histogram of the number of boundary pixels over the four types of local occlusion relations to classify among four types of object-container spatial relations: *O* is in front of *C*, *O* is behind *C*, *O* is inside *C* (and also occluded by *C*) as seen from a low view, or *O* in inside *C* (and is not occluded by *C*) as seen from a high view. In our experiments, we used a K-means clustering with K=4 to classify the spatial relations.